\documentstyle[preprint,aps]{revtex}
\begin{document}
\draft
\preprint{}
\title{General Relativistic Analog Solutions for Yang-Mills Theory
\thanks{For the memorial issue of {\it Theo. Math. Phys.}
in memory of F.A. Lunev}}
\author{D. Singleton}
\address{Department of Physics, California State University, Fresno, 
2345 East San Ramon Ave., Fresno, CA 93740-8031}
\date{\today}
\maketitle
\begin{abstract}
Finding solutions to non-linear field theories, such as
Yang-Mills theories or general relativity, is usually
difficult. The field equations of Yang-Mills theories and 
general relativity are known to share some mathematical
similarities, and this connection
can be used to find solutions to one theory using 
known solutions of the other theory. For example, the
Schwarzschild solution of general relativity can be shown to
have a mathematically similar counterpart in Yang-Mills
theory. In this article we will discuss several solutions 
to the Yang-Mills equations which can be found using this connection
between general relativity and Yang-Mills theory. Some
comments about the possible physical meaning of
these solutions will be discussed. In particular it will be argued
that some of these analog solutions of Yang-Mills theory may
have some connection with the confinement phenomenon. To this end
we will briefly look at the motion of test particles moving in the
background potential of the Schwarzschild analog solution.
\end{abstract}
\newpage
\narrowtext

\section{Introduction} 

Yang-Mills theories \cite{mills} are non-Abelian gauge theories 
which have found a central role in particle physics in describing
both the electroweak and strong interactions. The non-Abelian
nature of  Yang-Mills theories make the field equations
non-linear, and therefore much more difficult to handle
compared to Abelian gauge theories such as pure electromagnetism.
For example, at the classical level (and also approximately at the 
quantum level if the quantum corrections are not too large -- 
see the Introduction of Ref. \cite{jackson}) one can use 
superposition for Abelian gauge theories, while even at the classical  
level, superposition is not valid for Yang-Mills theories.
This non-linear nature of the Yang-Mills field 
equations makes finding solutions difficult. 
There are some well known and interesting solutions
to the Yang-Mills field equations, such as the t 'Hooft-Polyakov
monopole \cite{thooft}, the Julia-Zee dyon \cite{zee}, the
BPS dyon \cite{bogo} \cite{prasad}, and the instanton \cite{poly}, but
there is no systematic way of arriving at solutions to the Yang-Mills
field equations.

General relativity can also in some sense be considered a non-Abelian
gauge theory \cite{uti} \cite{carmeli}, and a mathematical connection
between the two theories can be made \cite{lun2} \cite{lunj}.  Using this
connection between the two theories one can ask if the solutions
to the field equations of one theory could provide a
starting point to look for solutions in the other theory. This is
in fact possible, and one can find a host of solutions in this
manner. In this paper we will give a review of the various solutions
found in this way and discuss some of their properties. All of the
solutions discovered in this way have the apparent weak point
that they have an infinite field energy, {\it i.e.}, 
there are singularities in the fields of the solutions 
which make the field energy infinite. This is to be contrasted with
the finite field energy solutions of Refs. \cite{thooft} 
\cite{zee} \cite{bogo} \cite{prasad}. However, aside from the
mathematical interest in studying all types of solutions that occur
in such non-linear field theories, we present some ideas concerning
the possible physical uses of such singular solutions. One speculation
is that some of these solutions may be connected with the confinement
phenomenon of the strong interaction. Just as the various black hole
solutions (Schwarzschild or Kerr black holes) exhibit a type of
confinement for any particle which crosses the event horizon, so too
the Yang-Mills analogs of these solutions may exhibit a
confining behaviour.

The outline of the paper is as follows : First we will discuss
the spherically symmetric solutions of the SU(2) Yang-Mills
equations coupled to a scalar field (these are usually called the
Yang-Mills-Higgs equations). Second we will discuss solutions
for gauge groups other than SU(2).
Finally we will examine the behaviour of a test particle which
is placed in the potential of the Schwarzschild analog solution. 
We will see that under certain conditions this analog
solution can confine the test particle, and that this system
has a half-integer angular momentum even though all the fields
involved are of integer angular momentum.

\section{SU(2) Yang-Mills Field Equations for Spherically Symmetric
Field Configurations}

The system studied in this section is an SU(2) gauge theory
coupled to a scalar field, $\phi ^a$, in the triplet representation.
The scalar field is taken to have no mass or self interaction.
The Lagrangian for this system is
\begin{equation} 
\label{lagrange}
{\cal L} = -{ 1\over 4} G_{\mu \nu} ^a G^{\mu \nu} _a +
{1 \over 2} (D_{\mu} \phi _a ) (D^{\mu} \phi ^a )
\end{equation}
where $G_{\mu \nu} ^a$ is the field strength tensor of the SU(2)
gauge fields, which is defined in terms of the gauge fields $W_{\mu} ^a$, 
as
\begin{equation} 
\label{fst}
G_{\mu \nu} ^a = \partial _{\mu} W_{\nu} ^a - \partial _{\nu} W_{\mu} ^a
+ g \epsilon _{abc} W_{\mu} ^b W_{\nu} ^c
\end{equation} 
and $D_{\mu}$ is the covariant derivative of the scalar field which is
given by
\begin{equation}
\label{codev}
D_{\mu} \phi ^a = \partial _{\mu} \phi ^a + g \epsilon _{abc} W_{\mu} ^b
\phi ^c
\end{equation}
The general equations of motion for this system are 
\begin{eqnarray}
\label{eqnmom}
\partial ^{\nu} G _{\mu \nu } ^a &=& g \epsilon _{abc} \left[
 G_{\mu \nu b} W^{\nu} _c - (D_{\mu} \phi _b) \phi _c \right] 
\nonumber \\
\partial ^{\mu} D_{\mu} \phi ^a &=& g \epsilon _{abc} (D_{\mu} \phi _b)
W_c ^{\mu}
\end{eqnarray}
these field equations can be simplified through the use of a generalized
Wu-Yang ansatz \cite{yang} which was used by Witten \cite{wit}
to study multi-instanton solutions
\begin{eqnarray}
\label{wuyang}
W_i ^a &=& \epsilon _{aij} {r^j \over g r^2} [1 - K(r)] + 
\left( {r_i r_a \over r^2} - \delta _{ia} \right)
{G(r) \over g r} \nonumber \\
W_0 ^a &=& {r^a \over g r^2} J(r) \nonumber \\
\phi ^a &=& {r^a \over g r^2} H(r)
\end{eqnarray}
$K(r)$, $G(r)$, $J(r)$, and $H(r)$ are the ansatz functions to be 
determined by the equations of motion. Inserting these 
expressions into the field equations in Eq. (\ref{eqnmom}) we
find the following set of coupled, non-linear equations, 
\begin{eqnarray}
\label{difeq}
r^2 K'' &=& K (K^2 + G^2 + H^2 - J^2 - 1) \nonumber \\
r^2 G'' &=& G (K^2 + G^2 + H^2 - J^2 -1) \nonumber \\
r^2 J'' &=& 2J (K^2 + G^2) \nonumber \\
r^2 H'' &=& 2 H (K^2 + G^2)
\end{eqnarray}
where the primes denote differentiation with respect to $r$. The
most well known solutions to these equations are those 
discovered by Prasad and Sommerfield \cite{prasad} and independently by 
Bogomolnyi \cite{bogo}. They are
\begin{eqnarray}
\label{soln}
K(r) &=& cos (\theta) C r \; csch(Cr) \; \; \; \; \; \; \; \; \; \; \; \;
G(r) = sin (\theta) C r \; csch(Cr) \nonumber \\
J(r) &=& sinh(\gamma ) [1 - C r \; coth(C r)] \; \; \; \; \; \; \; 
H(r) = cosh (\gamma ) [ 1 - C r \; coth(C r)]
\end{eqnarray}
where $C$, $\theta$ and $\gamma$ are 
arbitrary constants. One of the nice
properties of this solution is that it has finite field
energy. In terms of the ansatz functions the energy
density of the fields is
\begin{eqnarray}
\label{energy}
T^{00} &=& {1 \over g^2 r^2 } \Bigg( {K'}^2 + {G'}^2 + 
{(K^2 + G^2 - 1)^2 \over 2 r^2} + {J^2 (K^2 + G^2) \over r^2} 
+ {(rJ' - J)^2 \over 2 r^2} \nonumber \\
&+& {H^2 (K^2 + G^2) \over r^2}
+ {(r H' - H)^2 \over 2 r^2} \Bigg)
\end{eqnarray}
For the solution in Eq. (\ref{soln}) this gives a non-singular
energy density, which when integrated over all space yields
a finite field energy of $E = 4 \pi C cosh^2(\gamma) / g^2$.
This finite energy property  of the BPS solution is one of the
main reasons for the interest in this classical
solution. We now examine the general relativistic analog solutions
of the Yang-Mills equations.

\subsection{Solutions with Spherical Singularities}

To find the general relativistic analog solutions to the Yang-Mills
field equations we begin by examining the Schwarzschild solution
of general relativity.
The Schwarzschild solution in Schwarzschild coordinates has two
non-trivial components to the metric tensor : $g_{00}$ and $g_{rr}$.
The non-trivial spatial element has the form $g_{rr} =  {K r \over
1 - K r}$ and $g_{00} = - 1/ g_{rr}$ where $K = 1 / 2GM$. 
Trying this form of $g_{rr}$ in Eq. (\ref{difeq}) 
one immediately finds the following solution
\begin{eqnarray}
\label{schsol}
K(r) &=& {\mp \cos \theta C r \over 1 \pm C r} \; \; \; \;
G(r) = {\mp \sin \theta C r \over 1 \pm C r} \nonumber \\
J(r) &=& {\sinh \gamma \over 1 \pm C r} \; \; \; \; \; \; 
H(r) = {\cosh \gamma \over 1 \pm C r}
\end{eqnarray}
where $C, \gamma$ and $\theta$ are arbitrary constants. The solution
with the minus sign in the denominator (which we call the 
Schwarzschild-like solution) develops a singularity in the
gauge fields ($W_{\mu} ^a$) and scalar fields ($\phi ^a$) on a
spherical surface of radius $r = r_0 = 1/C$. Both the
Schwarzschild-like solution and the solution with the plus sign
in the denominator  
develop singularities in the fields at $r=0$. These field
singularities lead to the field energy of these solutions being
infinite, as can be seen by inserting the ansatz functions from
Eq. (\ref{schsol}) into Eq. (\ref{energy}) and trying to
integrate over all space. The investigation
of such infinite energy solutions to the Yang-Mills equations
has been discussed by several authors \cite{rosen} \cite{swank} \cite{proto}
\cite{maha} \cite{lunev} \cite{sing1}, and the earliest discussion 
actually pre-dates \cite{rosen} the study of the finite energy solutions
such as the t 'Hooft-Polyakov monopole or the BPS dyon. 

Although the infinite field energy of these solutions could
be seen as a drawback as compared to the finite energy solutions,
there are other classical field theory solutions which nevertheless
have a physical importance. The prime example is 
the Coulomb solution in electromagnetism which
has a field singularity at $r=0$ that is similar to the $r=0$
singularities of the solutions in Eq. (\ref{schsol}). 
The Schwarzschild-like solution, with its singular spherical
surface, has been speculated to have some connection with the
confinement phenomenon for quarks \cite{swank} \cite{maha} 
\cite{lunev} \cite{yoshida} \cite{pav}. By studying the motion of a test 
particle which moves in the potentials given by the minus
sign solution in Eq. (\ref{schsol}) one finds that the spherical
singularity in the fields represents a barrier which can trap
the test particle inside the sphere. This is similar in spirit
to bag models of hadron structure where one looks at test
particles moving in some confining potential (such as an
infinite spherical well). Also it is interesting that 
this Schwarzschild-like solution was arrived at from the
general relativistic solution for a non-rotating black hole,
which exhibits its own type of ``confinement'' : any particle
which passes within the event horizon becomes permanently trapped.
One should be cautious about pushing this analogy too far, since
the nature of the spherical singularity in general relativity and Yang-Mills
theory are different. The singularity at the event horizon of
the general relativistic Schwarzschild solution is not a physical
singularity, but a coordinate singularity as can be seen by 
writing the Schwarzschild solution in Kruskal coordinates, where
the only singularity is the one at $r=0$. Both singularities in
the Yang-Mills analog of the Schwarzschild solution are true
singularities in the fields. 

The existence of singular solutions for certain field theories is
not new ({\it e.g.} the singularities in the Coulomb
solution of electromagnetism, the Wu-Yang monopole solution
\cite{yang}, or the meron solutions \cite{dea}).  Even
the appearance of a singularity in the gauge fields
on a spherical surface, such as occurs in the Schwarzschild-like
solution of Eq. (\ref{schsol}), which may at first appear unique, 
can be found in other infinite energy solutions. These other
solutions  possess an infinite set of concentric
spherical surfaces on which the fields develop a singularity. This
could be taken as evidence that such spherical surfaces with
singularities are not uncommon features in classical solutions to
the Yang-Mills field equations. The first of these solutions can be
obtained by exchanging the hyperbolic functions of the BPS 
solution in Eq. (\ref{soln}) with their trigonometric counterparts
\begin{eqnarray}
\label{soln1}
K(r) &=& cos( \theta ) C r \; csc(Cr) \; \; \; \; \; \; \; \; \; \; \; 
\; G(r) = sin ( \theta ) C r \; csc(Cr) \nonumber \\
J(r) &=& sinh(\gamma ) [1 - C r \; cot(C r)] \; \; \; \; \; \; \;
H(r) = cosh (\gamma ) [ 1 - C r \; cot(C r)]
\end{eqnarray}
This solution was briefly discussed by Hsu and Mac \cite{hsu} in
their derivation of the BPS solution ({\it i.e.} Hsu and
Mac start with a solution like that in Eq. (\ref{soln1}) and
apply the transformation $C \rightarrow i C$ to arrive at the
BPS solution). This solution exhibits a series of concentric
spherical surfaces on which the gauge and scalar fields become
singular. These singularities are located on the spherical surfaces
$C r = n \pi $ where $n =1,2,3,4 ...$. Inserting the ansatz
functions of Eq. (\ref{soln1}) in Eq. (\ref{energy}) we find
that the energy density of this solution is
\begin{equation}
\label{ensoln1}
T^{00} = {2 cosh ^2 (\gamma) \over r^2 g^2} \left[ C^2 csc ^2 (C r)
\Big( 1 - C r \; cot (C r)\Big) ^2 + {\Big( C^2 r^2 
csc ^2 (C r) - 1 \Big) ^2 \over 2 r^2} \right]
\end{equation}
The energy density becomes singular on the same spherical surfaces 
as the gauge and scalar fields. These spherical shells, on which
the energy density becomes infinite, cause the total field
energy of this solution to diverge.

To obtain the next solution we simply try the complementary
trigonometric functions for the solution in Eq. (\ref{soln1}).
Doing this shows that the following is also a solution \cite{sing3} 
to the Eq.(\ref{difeq})
\begin{eqnarray}
\label{soln2}
K(r) &=& cos( \theta ) C r \; sec(Cr) \; \; \; \; \; \; \; \; \; \; \; \;
G(r) = sin( \theta ) C r \; sec(Cr) \nonumber \\
J(r) &=& sinh(\gamma ) [1 + C r \; tan(C r)] \; \; \; \; \; \; \;
H(r) = cosh (\gamma ) [ 1 + C r \; tan(C r)]
\end{eqnarray}
It should be noted that due to the linear $C r$ term in each solution,
one can not obtain the solution in Eq. (\ref{soln2}) from  
the other trigonometric solution in Eq. (\ref{soln1}) by 
simply letting $C r \rightarrow C r - \pi /2$. Although 
these two trigonometric solutions are in this sense 
distinct ({\it i.e.} they are not simply related by the
transformation $C r \rightarrow C r - \pi / 2$) they
are physically similar since most of the comments concerning 
the solution in Eq. (\ref{soln1}) apply here as well.
Most obviously the ansatz functions, and therefore the gauge
and scalar fields, become singular when $C r = n \pi / 2$
where $n = 1, 3, 5, 7, ... \;$ and at $r=0$. 
Thus this solution exhibits a series
of concentric spherical surfaces on which its fields become
singular as well as a point singularity  at the origin. 
These singularities also show up in the energy density
of this solution as they did for the solution in Eq. (\ref{soln1}). 
The point singularity at $r=0$ and the spherical
singular surfaces of the solutions in Eqs. (\ref{soln1})
(\ref{soln2}) are similar to that of the 
solutions from Eq. (\ref{schsol}). However, the solutions in Eq. 
(\ref{schsol}) only possessed one spherical surface 
on which the fields and energy density diverged. One conjectured 
use for the Schwarzschild-like solution is as a
possible explanation of the confinement mechanism. When the 
Schwarzschild-like solution of Ref. \cite{sing1} is treated as a
background potential in which a test particle is placed it is
found that the spherical singularity can act as an impenetrable
barrier which traps the test particle either in the interior
or the exterior of the sphere \cite{yoshida}, giving a classical
type of confinement. Similar results have been found 
for other singular solutions \cite{swank} \cite{maha} \cite{lunev}.
In addition Ref. \cite{swank} points out that such a classical
type of confinement is only possible with infinite energy solutions.
Treating the trigonometric solutions as a background potential would also trap
test particles between any two of the concentric spherical singularities.
These trigonometric solutions could possibly be used to solve 
the field equations in some limited range of $r$, and then 
it could be patched to one of the other solutions which 
would solve the field equations for the remaining range of $r$.
This is similar to what is sometimes done in general relativity
where one tries to patch an exterior solution 
with some interior solution.

Finally one can obtain a third solution to Eq. (\ref{difeq}) by
applying the transformation $C \rightarrow i C$ to the
solution \cite{sing3} in Eq. (\ref{soln2}). This yields
\begin{eqnarray}
\label{soln3}
K(r) &=& {\bf i} cos(\theta ) C r \; sech(Cr) \; \; \; \; \; \; \; \; \; 
\; \; \; G(r) = {\bf i} sin(\theta ) C r \; sech(Cr) \nonumber \\
J(r) &=& sinh(\gamma ) [1 - C r \; tanh(C r)] \; \; \; \; \; \; \;
H(r) = cosh (\gamma ) [ 1 - C r \; tanh(C r)]
\end{eqnarray}
Since the ansatz functions $K(r)$ and $G(r)$ are imaginary, the 
space components of the gauge fields will be complex. Despite this
all the physical quantities associated with this complex solution, 
such as energy density, are real. Inserting the ansatz functions of Eq. 
(\ref{soln3}) into Eq. (\ref{energy}) we find that the field 
energy density is
\begin{equation}
\label{ensoln2}
T^{00} = {2 cosh ^2 (\gamma) \over r^2 g^2} \left[ - C^2 sech ^2 (C r)
\Big( 1 - C r \; tanh(C r) \Big) ^2 + {(C^2 r^2 sech ^2 (C r) + 1)^2
\over 2 r^2} \right]
\end{equation}
This energy density is real, but the total field energy 
is infinite due to the singularity
at $r = 0$. Thus the above solution is more like a Wu-Yang
monopole \cite{yang} or a charged point particle, as opposed
to a finite energy BPS dyon. 

\subsection{SU(2) Solutions With Increasing Potentials}

In addition to the preceding infinite energy solutions which have
gauge and scalar fields that become singular on some spherical surface,
there are other types of infinite energy, general relativistic 
analog solutions. In general relativity if one allows 
for a nonzero cosmological constant,
$\Lambda$, then the time-time component of the metric tensor for
the Schwarzschild solution becomes \cite{ohanian}
\begin{equation}
\label{cossch}
g_{00} = 1 - {2 G M \over r} - {\Lambda r^2 \over 3}
\end{equation}
The Newtonian potential for this solution is
\begin{equation}
\label{newsol}
\Phi = {( g_{00} - 1) \over 2} = {-G M \over r} - {\Lambda r^2 \over 6}
\end{equation}
Using Eq. (\ref{newsol}) as a starting point one finds the following
simple solution \cite{sing2} to Eq. (\ref{difeq})
\begin{equation}
\label{linear}
K(r) = cos \theta \; \; \; \; \; G(r) = sin \theta \; \; \; \; \;
J(r) = H(r) = { B \over r} + A r^2
\end{equation}
where $a, B$ and $\theta$ are arbitrary constants. If one sets
$A = 0$ then it can be seen the Schwarzschild-like solutions
of Eq. (\ref{schsol}) and those above in Eq. (\ref{linear}) are
of a similar form in the limit $C \rightarrow \infty$
and $e^{\gamma} / C \rightarrow 2 B$. Inserting the ansatz
functions of Eq. (\ref{linear}) into the gauge and scalar fields 
of Eq. (\ref{wuyang}) one finds that the time component of the gauge
field ($W_0 ^a$) and the scalar field ($\phi ^a$) behave like 
$A r + B/ r^2$. The space part of the gauge fields 
($W_i ^a$) have a $1/r$ dependence. This classical
solution exhibits a linear confining potential similar to those
used in some phenomenological studies of hadronic spectra \cite{eich}.
In addition lattice gauge theory arguments \cite{wilson} seem to indicate 
that the confining potential between quarks should be linear. Classical
solutions similar to those in Eq. (\ref{linear}) were also discussed
in Ref. \cite{sivers} in connection with the confinement problem.

This solution also has an infinite field energy. Inserting the
ansatz functions of Eq. (\ref{linear}) into the energy density
of Eq. (\ref{energy}) and integrating to get the total field
energy one finds
\begin{eqnarray}
\label{enlin}
E &=& \int T^{00} d^3 x = { 4 \pi \over g^2} \int _{r_a} ^{r_b} T^{00}
r^2 dr \nonumber \\
&=& {4 \pi A^2 \over g^2} (r_b ^3 - r_a ^3) - {8 \pi B^2 \over g^2}
\left( {1 \over r_b^3} - {1 \over r_a ^3} \right)
\end{eqnarray}
where we have introduced an upper ($r_b$) and lower ($r_a$) 
cutoff in the radial coordinate. If one lets $r_b \rightarrow
\infty$ then the field energy becomes infinite due to the linear
part of the gauge and scalar fields, while if one lets $r_a 
\rightarrow 0$ then the field energy becomes infinite due to
the singularity at $r=0$. Compared to the solutions in Eq.
(\ref{schsol}), which had infinite field energy from local 
singularities (either at $r=0$ or $r= 1/C$), the solution
in Eq. (\ref{linear}) can have a infinite field energy from
the point singularity at $r=0$ and/or the linearly increasing gauge
and scalar fields as $r \rightarrow \infty$. Again, although this
classical solution has some undesirable characteristics, it also
exhibits features which are found in some phenomenological
studies of hadronic bound states.

\subsection{SU(3) Solutions}

Up to this point we have discussed classical solutions to
the Yang-Mills field equations for SU(2) fields. Since QCD
involves the SU(3) gauge group it is natural to ask if
there are any SU(3) or even SU(N) generalizations of the
above solutions. One possibility is to embed the above SU(2)
solutions into an SU(N) gauge theory \cite{sing4}. Recently
\cite{dzhunu} a Schwarzschild-like classical solution was found
which is not a simple embedding of the previous SU(2) solutions
into an SU(N) gauge theory, but is a true SU(3) solution. To 
arrive at the SU(3) solution one makes the
following generalization \cite{dzhunu} of the Wu-Wu ansatz 
\cite{wuwu} \cite{pagel} \cite{volkov}
\begin{eqnarray}
\label{wuwu}
W _0 &=& {- {\bf i} \phi (r) \over g r^2} \left( \lambda ^7 x
-\lambda ^5 y + \lambda ^2 z \right) + {1 \over 2} \lambda ^a
\left( \lambda ^a _{ij} + \lambda ^a _{ji} \right) {x^i x^j \over
g r^3} w(r) \nonumber \\
W^a _i &=& \left( \lambda ^a _{ij} - \lambda ^a _{ji} \right)
{{\bf i} x^j \over g r^2} (1 - f(r)) + \lambda ^a _{jk} \left(
\epsilon _{ilj} x^k + \epsilon_{ilk} x^j \right) {x^l \over g r^3}
v(r)
\end{eqnarray}
where $\lambda ^a$ are the Gell-Mann matrices. Using this ansatz
in the Yang-Mills field equations yields the following set
of coupled differential equations for the functions $f(r) , v(r) ,
\phi (r)$ and $w(r)$
\begin{eqnarray}
\label{difeqsu3}
r^2 f'' &=& f^3 - f + 7 f v^2 + 2vw\phi - f (w^2 + \phi ^2)
\nonumber \\
r^2 v'' &=& v^3 - v + 7v f^2 + 2 f w \phi - v(w^2 + \phi ^2 )
\nonumber \\
r^2 w'' &=&  6 w (f^2 + v^2) - 12 f v \phi \nonumber \\
r^2 \phi '' &=& 2 \phi (f^2 + v^2) - 4 f v w
\end{eqnarray}
where the primes denote differentiation with respect to $r$.
The nonlinear, coupled differential equations in Eq. (\ref{difeqsu3})
are the SU(3) equivalents of the equations in Eq. (\ref{difeq}).
In Ref. \cite{dzhunu} several simplifying assumptions were made
in order to make the problem more tractable. First, taking
$w = \phi = 0$, reduces Eq. (\ref{difeqsu3}) to
\begin{eqnarray}
\label{su3bag}
r^2 f'' &=& f (f^2 - 1 + 7 v^2) \nonumber \\
r^2 v'' &=& v(v^2 - 1 + 7 f^2)
\end{eqnarray}
Then further simplifying by letting $f^2 = v^2 = q^2 / 8$ one
finds that Eq. (\ref{su3bag}) reduces to the Wu-Yang \cite{yang}
equation for $q(r)$ 
\begin{equation}
\label{su3bag1}
r^2 q'' = q( q^2 - 1)
\end{equation}
This equation has been shown \cite{rosen} \cite{lunev} \cite{dzhunu}  
to have a solution which is singular at some radius $r = r_1$. In
other words near $r = r_1$ the solution will be of the form 
\begin{equation}
\label{dzhsol}
q(r) \approx {A \over r_1 - r}
\end{equation}
where $A$ and $r_1$ are constant. Thus, even with the scalar field
absent one can find solutions to the pure gauge field theory 
equations which will tend to trap test particles behind a spherical
barrier in much the same way as the Schwarzschild-like solution
of Eq. (\ref{schsol}). It is also possible to find closed form
solutions to a special case of the system in Eq. (\ref{difeqsu3})
Taking $v=w=0$ then the equations of Eq. (\ref{difeqsu3}) become
\begin{eqnarray}
\label{su3bag2}
r^2 f'' &=& f(f^2 - \phi ^2 -1) \nonumber \\
r^2 \phi '' &=& 2 \phi f^2
\end{eqnarray}
which has the following simple closed form solution
\begin{eqnarray}
\label{mysu3}
f(r) &=& \mp {C r \over 1 \pm C r} \nonumber \\
\phi (r) &=& \pm {i \over 1 \pm C r}
\end{eqnarray}
Other, similar solutions can be found by making different
simplifying assumptions such as $f=w=0$. 
Thus solutions with singular fields on a spherical surface are not
unique to SU(2) gauge theories, but can also be found for SU(3)
\cite{dzhunu} and in general for SU(N) \cite{sing4}. The interesting
aspect of the solutions given by Dzhunushaliev in Ref. \cite{dzhunu}
is that these solutions are true SU(3) solutions rather than
embeddings of the SU(2) solution into the SU(N) gauge group as
in Ref. \cite{sing4}. Also the SU(3) solutions presented here
are pure gauge field solutions, as opposed to the general SU(2) 
solutions for the system given in Eq. (\ref{lagrange}), which involves
scalar fields. In some sense the role of the scalar field
of the SU(2) system is taken up by the time component of the gauge
field in the SU(3) system. This can be seen by comparing the system
of equations of Eq. (\ref{difeq}) with the system of equations of
Eq. (\ref{difeqsu3}) : the equations for $f(r), v(r)$ are
similar to those for $K(r), G(r)$ while the equations for
$w(r), \phi (r)$ are similar to those for $J(r), H(r)$. 

\section{Electromagnetic Properties of the SU(2) Solutions}

All of the SU(2) solutions to the Yang-Mills field equations
have interesting ``electromagnetic'' features.
To investigate these properties we will use 't Hooft's 
definition of a generalized, gauge invariant, U(1) field 
strength tensor \cite{thooft}
\begin{equation}
\label{emfst2}
F_{\mu \nu} = \partial _{\mu} (\hat{ \phi} ^a W^a _{\nu}) -
\partial _{\nu} (\hat{ \phi} ^a W^a _{\mu}) - {1 \over g}
\epsilon _{abc} \hat{\phi } ^a ( \partial _{\mu} \hat{\phi } ^b )
( \partial _{\nu} \hat{\phi } ^c )
\end{equation}
where $\hat{\phi } ^a = \phi ^a (\phi ^b \phi ^b)^{-1/2}$. This
generalized U(1) field strength tensor reduces to the usual
expression for the field strength tensor if one performs a
gauge transformation to the Abelian gauge where the scalar
field only points in one direction in isospin space ({\it i.e.}
$\phi ^a = \delta ^{3a} v$) \cite{arafune}. If one associates
this U(1) field with the photon of electromagnetism then the
solutions in Eqs. (\ref{schsol}) (\ref{soln1}) (\ref{soln2})
(\ref{soln3}) (\ref{linear}) carry magnetic and/or electric charges. 
In general the electric and magnetic fields associated with these 
solutions are
\begin{eqnarray}
\label{ebschw}
E_i &=& F_{i0} = {r_i \over g r} {d \over dr} \Bigg( {J(r)
\over r} \Bigg)   \nonumber \\
B_i &=& {1 \over 2} \epsilon _{ijk} F_{jk} = -{r_i \over
g r^3}
\end{eqnarray}
The magnetic field of all the solutions is that of a point monopole 
of strength $-4 \pi / g$. The reason for this will be discussed
at the end of this section.

The electric field of the Schwarzschild-like solutions of Eq.
(\ref{schsol}) is easily found by inserting the ansatz function
$J(r)$ from Eq. (\ref{schsol}) into Eq. (\ref{ebschw}). Doing this
gives
\begin{equation}
\label{eb}
E_i = {- r_i sinh \gamma (1 \pm 2 C r) \over g r^3 (1 \pm C r) ^2}
\end{equation}
As $r \rightarrow \infty$ this electric field goes as $1 / r^3$
which indicates that the net electric charge of this solution
is zero, although there appears to be some kind of dipole charge
distribution.

The electric fields of both of the trigonometric solutions presented
in Eqs. (\ref{soln1}) (\ref{soln2}) are similar to each other in
that they indicate that these solutions carry an infinite electric
charge. Taking the ansatz function $J(r)$ from the trigonometric
solution given in Eq. (\ref{soln1}) and inserting it into Eq. 
(\ref{ebschw}) yields the following electric field
\begin{equation}
\label{eb2}
E_i = { - sinh (\gamma) r_i \over g r^3}
\big( 1 - C^2 r^2 csc ^2 (C r) \big)
\end{equation}
The electric field does not fall off for large $r$, but 
behaves like $r_i csc ^2 (C r) / r$. This electric field
also becomes singular on the spherical surfaces defined by
$C r = n \pi$ where $n = 1,2,3,4 ...$. The trigonometric
solution of Eq. (\ref{soln2}) exhibits the same type of electric
field except that it becomes
singular on the spherical shells given by $C r = n \pi /2$
(with $n = odd$) and at $r=0$.
The electric charge of this solution is also infinite 
since the electric field from Eq. (\ref{eb2}) does not
fall off as $r \rightarrow \infty$. For the special case
where $\gamma = 0$ , one finds that the solution carries
no electric charge, but only a magnetic charge. Even in this
case the energy density becomes singular on the
concentric spherical surfaces and at the origin. 
Both the BPS solution and the solutions from Eqs. (\ref{soln1}) 
(\ref{soln2}) have the same finite magnetic strength of
$-4 \pi / g$. Although this solution is a
dyon in the sense that it carries both magnetic and electric charge
it is probably not correct to view it as a particle-like
solution, since the electric field does not fall off, thus
implying that these solutions have an infinite, spread out
electric charge.

The electric field associated with the
complex solution in Eq. (\ref{soln3}) can be found in the same way
as for the other solutions. Inserting the ansatz function $J(r)$
from Eq. (\ref{soln3}) into Eq. (\ref{ebschw}) yields
\begin{equation}
\label{eb3}
E_i = { - sinh (\gamma) r_i \over g r^3}
\Big( C^2 r^2 sech ^2 (C r) + 1 \Big)
\end{equation}
As with all the other solutions, the complex solution carries a magnetic
charge of strength $-4 \pi / g$. In addition, by examining the behaviour
of the electric field in Eq. (\ref{eb3}) as $r \rightarrow \infty$
one finds that this complex solution carries an electric charge of
$-4 \pi sinh (\gamma )/ g$, which is the same as that carried by the BPS
solution. One interesting feature
of the solution from Eq. (\ref{soln3}) is that
even though the space components of the gauge fields are complex
all the physical quantities ({\it e.g.} field energy, magnetic
charge, electric charge) calculated from it are real. Also, unlike the 
solutions of Eqs. (\ref{soln1}) (\ref{soln2}), this complex
solution can be viewed as a point-like dyon since it has a localized
electric charge. The main difference between this solution and 
the BPS solution is the infinite field energy of the complex 
solution due to the field singularity at $r=0$.

While many of the physical characteristics of these various 
solutions are substantially different in each case, the
magnetic charge of all the solutions is the same.
This comes about since the magnetic charge of each solution is
a topological charge which carries the same value for each field 
configuration. A topological current, $k _{\mu}$, can be
defined as \cite{arafune}
\begin{equation}
\label{current}
k_{\mu} = {1 \over 8 \pi} \epsilon_{\mu \nu \alpha \beta} 
\epsilon _{abc} \partial ^{\nu} {\hat \phi} ^a \partial ^{\alpha} 
{\hat \phi} ^b \partial ^{\beta} {\hat \phi} ^c
\end{equation}
The topological charge of this field configuration is then
\begin{eqnarray}
\label{tc}
q &=& \int k_0 d^3 x  = {1 \over 8 \pi} \int 
(\epsilon_{ijk} \epsilon_{abc}
\partial ^{i} {\hat \phi} ^a \partial ^{j} {\hat \phi} ^b
\partial ^{k} {\hat \phi} ^c ) d^3 x \nonumber \\
&=& {1 \over 8 \pi} \int \epsilon_{ijk} \epsilon_{abc}
\partial ^{i} ( {\hat \phi} ^a \partial ^{j} {\hat \phi} ^b
\partial ^{k} {\hat \phi} ^c ) d^3 x
\end{eqnarray}
For all the solutions one finds that ${\hat \phi } ^a = r^a / r$, 
which is the same regardless of the ansatz function $H(r)$. In all
cases we find that the topological charge is $q =1$. In the next
section when we examine the motion of a test particle in the background
field of the Schwarzschild-like solution we will find that there
is a field angular momentum due to the interaction of the test
particle with the field configuration of the Schwarzschild-like
solution. This field angular momentum can be seen to arise from the 
interaction of the topological magnetic charge with the charge
of the test particle, in much the same way that the configuration of
a normal magnetic charge and an electric charge lead to a field 
angular momentum \cite{thomson} \cite{saha} \cite{jackson}.

\section{Motion of Tests Particles in Schwarzschild-like Potential}

We would now like to study the motion of a test particle in the
background potential of the Schwarzschild-like solution of Eq.
(\ref{schsol}). We will make several assumptions in doing this.
First we will take our test particle to be a scalar particle
as in Ref. \cite{lunev} \cite{yoshida}. One reason for making this
choice is to illustrate the spin from isospin \cite{rebbi} effect
that occurs with these solutions. As discussed in the preceding section
all of these solutions carry a magnetic charge. Many researchers have
remarked on the fact \cite{thomson} \cite{saha} that the composite
system of a magnetic charge and electric charge carry an angular
momentum due to the configuration of electric and magnetic fields.
Even when the magnetic charge is topological in character, as is the
case with 't Hooft-Polyakov monopoles, one finds \cite{rebbi} a similar
effect whereby the composite system of a topological magnetic charge
and a particle with the charge of the gauge group, will carry an
angular momentum in the gauge fields. This has the interesting
consequence that if one wants to construct fermionic objects from
the singular solutions one should use scalar particle which are
in the fundamental representation of the gauge group -- SU(2) for the
solutions considered here. (Fermionic test particles in the adjoint 
representation would also give a net fermionic bound state \cite{pav}).
Our second assumption is that the test particle will be coupled to the
scalar field part of the solution of Eq. (\ref{schsol}) via the
following substitution $m^2 \rightarrow (m + \lambda \sigma ^a \phi ^a
/ 2)^2$ where $\lambda$ is an arbitrary coupling constant. Our final
assumption is to set $\theta = 0$ in Eq. (\ref{schsol}) in order to
not have to take the ansatz function $G(r)$ into account. In this way
the scalar particle $\Phi ^A$ moving in the background field of the
Schwarzshild-like solution given in Eq. (\ref{schsol}) becomes
\begin{eqnarray}
\label{kg1}
& &\left( \partial _0 - {i g \over 2} \sigma ^a W_0 ^a \right)
\left( \partial _0 - {i g \over 2} \sigma ^a W_0 ^a \right)^A _B
\Phi ^B  (x ,t) - \left( \partial _i - {i g \over 2} \sigma ^a W^a _i \right)
\left( \partial _i - {i g \over 2} \sigma ^a W^a _i \right) ^A _B
\Phi ^B (x , t) \nonumber \\
&=& - \left( m + {\lambda \over 2} \sigma ^a \phi^a
\right) ^2 \Phi ^A (x , t)
\end{eqnarray}
Where $\sigma ^a$ are the standard Pauli matrices, and $A, B$ are SU(2)
group indices which can take on the values $1$ or $2$.
Taking $\Phi ^A (x,t) = \Phi ^A (x) e^{-i E t}$, using $(\sigma ^a v^a) ^2 =
v^a v^a$ and expanding we find that Eq. (\ref{kg1}) becomes
\begin{eqnarray}
\label{kg2}
& &\left[- E^2 -g \sigma ^a W_0 ^a E - {g^2 \over 4} (W_0^a)^2 
- \nabla ^2 + i g \sigma ^a W_i^a \partial _i + {g^2 \over 4}
(W_i ^a)^2 \right] ^A _B \Phi ^B (x) \nonumber \\
= &-& \left[m^2 + \lambda m \sigma ^a \phi ^a
+ {\lambda ^2 \over 4} (\phi ^a)^2 \right] ^A _B \Phi^B (x)
\end{eqnarray}
Inserting the ansatz form of the gauge and scalar fields from Eq.
(\ref{wuyang}) into Eq. (\ref{kg2}) then yields
\begin{eqnarray}
\label{kg3}
& & - \left[\nabla ^2 - {(1-K(r) ) \over r^2} \sigma ^a l^a -
{(1-K(r))^2 \over 2 r^2} + {\sigma ^a r^a \over r^2} E J(r)
+ {J(r) ^2 \over 4 r^2} \right] ^A _B \Phi ^B (x) \nonumber \\
= & & \left( E^2 - m^2 - {\lambda m \over g r^2} \sigma ^a r ^a
H(r) - {\lambda ^2 \over 4 g^2 r^2} H(r) ^2 \right) ^A _B \Phi ^B (x)
\end{eqnarray}
where we have used $i g \sigma ^a W_i ^a \partial _i = -i (1 - K(r))
\sigma ^a \epsilon _{aji} r^j \partial ^i / r^2 = (1-K(r)) \sigma ^a
l^a / r^2$ ($l^a$ is the standard orbital angular momentum
operator), and $(W^a _i)^2 = \epsilon_{aij} \epsilon_{aik} r^j r^k
(1-K(r))^2 / g^2 r^4 = 2 (1-K(r))^2 / g^2 r^2$. Since $\sigma ^a
l^a$ does not commute with $\sigma ^a r^a$ Eq. (\ref{kg3}) is
difficult to handle. By taking advantage of the free parameter
$\gamma$ which occurs in the ansatz functions $J(r), H(r)$ one can chose
$\gamma$ such that $E sinh \gamma = \lambda m cosh \gamma / g$. With
this choice the two $\sigma ^a r^a$ terms in Eq. (\ref{kg3})
cancel one another. In order to handle the $\sigma ^a l^a$ term
it is necessary to define the total angular momentum operator as
\begin{equation}
\label{jtot}
J^a = l^a + S^a = l^a + {1 \over 2} \sigma ^a
\end{equation}
Thus the total angular momentum comes not only from the orbital
angular momentum, but has a contribution that looks like a spin
angular momentum. The $\sigma$ matrices in the last term of
Eq. (\ref{jtot}) are, however, connected with the isospin of the
system rather than with spin. This is just the spin from isospin
effect \cite{rebbi}, and is connected with the fact that the 
Schwarzschild-like solution of Eq. (\ref{schsol}) carries a topological
magnetic charge. Thus even though our system involves only integer
spin fields ({\it i.e.} $W_{\mu} ^a  , \phi ^a , \Phi ^A$) the
combined system is a spin 1/2 object. Using Eq. (\ref{jtot})
the $\sigma ^a l^a$ term can be expanded in the usual way as $\sigma ^a l^a 
= 2 S^a l^a = J_{op} ^2 -l _{op} ^2 -S _{op} ^2$ except
now $S _{op}$ is the isospin operator rather than the spin operator.
Finally, we make the simplifying assumption that $\lambda = g$ so that 
the $J(r)^2$ and $H(r) ^2$ terms may be more easily combined. At this 
point there seems to be nothing special in this choice, but we
will see that according to the arguments of Ref. \cite{lunev} and
\cite{exner}, the barrier at $r = 1/C$ will only absolutely confine
a particle if $\lambda \ge g$ while for $\lambda < g$ there will
be some probability for the test particle to tunnel through the
barrier. Combining all the preceding assumptions we find
\begin{eqnarray}
\label{kg4}
& & - \left[ \nabla ^2 - {(1-K(r)) \over r^2}(J_{op}^2 -l_{op}^2-S_{op}^2) -
{(1-K(r))^2 \over 2 r^2}+ {1 \over 4 r^2}(J(r)^2 - H(r)^2)
\right] ^A _B \Phi ^B (x) \nonumber \\
= & & \left( E^2 - m^2 \right) ^A _B \Phi ^B (x)
\end{eqnarray}
In order to separate out the radial equation from Eq. (\ref{kg4})
we take 
\begin{equation}
\label{sep}
\Phi ^A (x) = {1 \over r} f_{Jl} (r) Y^A _{JlM} (\theta , \phi )
\end{equation}
where the $Y^A _{JlM}$ are the standard spinor
spherical harmonics that one gets from adding an orbital angular
momenta $l^a$ to a spin $1/2$. Here spin is replaced by isospin, but
the math, and the spinor spherical harmonics, are exactly the same. 
Now inserting Eq. (\ref{sep}) into Eq. (\ref{kg4}) yields
\begin{eqnarray}
\label{kg5}
& &-\left[ {d^2 \over dr^2} - {D \over r^2} -{F(1-K(r)) \over r^2}
- {(1-K(r))^2 \over 2 r^2} + {1 \over 4 r^2}
(J (r)^2 - H (r)^2) \right] f_{Jl} (r) \nonumber \\
& & = (E^2 - m^2) f_{Jl} (r)
\end{eqnarray}
where we have defined the constants $D= l(l+1)$ , $F = J(J+1)-l(l+1)-3/4$. 
Then setting $x = C r$ and inserting the ansatz functions 
$K(r), J(r), H(r)$ from Eq. (\ref{schsol}) into Eq. (\ref{kg5})
turns the problem into an effective one-dimensional Schr{\"o}dinger
equation
\begin{equation}
\label{kg6}
\left[- {d^2 \over dx^2} + { D \over x^2} + {F (1-2x) \over x^2
(1-x)} + {(1-2x)^2 \over 2 x^2 (1-x)^2} +  {1 \over 4 x^2 (1-x)^2}
\right] f_{Jl} (x) = {(E^2 - m^2) \over C^2} f_{Jl} (x)
\end{equation}
where all the non-derivative terms on the left hand side are the
effective potential. The key feature of this effective 
potential are the singularities at $x=0$ and $x=1$. Now as
$x \rightarrow 1$ the leading term in the effective potential goes like
\begin{equation}
\label{effv}
V_{eff} (x) = {D \over x^2} + {F (1-2x) \over x^2 (1-x)}
+{(1-2x)^2 \over 2 x^2 (1-x)^2} +{1 \over 4 x^2 (1-x)^2}
\rightarrow {3 \over 4 (1-x)^2}
\end{equation}
It was argued in Refs. \cite{exner} \cite{lunev} that such
a singularity would only present a true barrier to the test
particle ({\it i.e.} the probability of the test particle tunneling
through the barrier would be zero) if the coefficient in Eq.
(\ref{effv}) were greater than or equal to $3/4$. Thus the effective
potential of Eq. (\ref{kg6}) just confines the test particle to remain
in the range $0\le x \le 1$. The fact that the effective potential
is just able to confine the test particle stems from our
choice of $\lambda = g$ for the coupling of the scalar potential
$\phi ^a$ to the test particle $\Phi ^A$. If we had taken $\lambda
< g$ then the coefficient in the limiting form of the effective
potential from Eq. (\ref{effv}) would have been less than $3/4$
and the test particle would no longer be confined ({\it e.g.} if
one took $\lambda =0$ it is straightforward, starting from Eq.
(\ref{kg3}), to show that one gets a coefficient of $1/2$). Conversely,
when $\lambda > g$ then the coefficient in Eq. (\ref{effv}) becomes
greater than $3/4$ and the test particle becomes confined. This has
the interesting implication that the scalar potential plays an
important role in this confinement mechanism. Although, generally
confinement is thought to be just the result of the gauge interaction,
there are phenomenological studies \cite{goebel} \cite{tekuda}
\cite{ram} which indicate 
that an effective scalar potential is involved in the confinement
mechanism.

To get more detailed in the solution of Eq. (\ref{kg6}) one must
pick particular values of $J$ and $l$ (which determine the constants
$D$ and $F$ in Eq. (\ref{kg6})), and solve for the eigenfunctions,
$f_{Jl} (x)$ and eigenvalues $(E^2 -m^2) /C^2$. In general this must be done
numerically \cite{pav} \cite{yoshida}, however, the key features of the
effective one-dimensional potential of Eq. (\ref{effv}) ({\it i.e.}
the singularities in the potential at $x=0$ and $x= 1$) make this
potential similar to the P{\"o}schl-Teller potential \cite{flugge}.
\begin{equation}
\label{postel}
V(x) = {1\over 2} V_0 \left[ {\alpha (\alpha -1) \over sin ^2 (\pi x /
2)} + {\beta (\beta -1) \over cos ^2 (\pi x / 2)} \right]
\end{equation}
where $\alpha , \beta , V_0$ are constants. By choosing $\alpha , \beta$
and $V_0$ correctly the P{\"o}schl-Teller potential can be made similar to
the effective potential from Eq. (\ref{effv}). Then the known
eigenfunctions and eigenvalues of the P{\"o}schl-Teller potential
should give a good approximation to the eigenvalues and eigenfunctions
of the potential from Eq. (\ref{effv}). The eigenfunctions for the
P{\"o}schl-Teller potential are \cite{flugge}
\begin{equation}
\label{ef}
f_n (x) = K sin ^{\alpha} (\pi x /2) cos ^{\beta} (\pi x / 2)
\; \; \; _2 F _1 \left( -n , \alpha + \beta +n , \alpha +{1 \over 2} ;
sin ^2 (\pi x / 2) \right)
\end{equation}
where $K$ is a constant fixed by normalization, $n$ is the radial
quantum number which takes on values of $n=0, 1, 2, 3 ...$, 
and $_2 F _1 (a,b,c ;x)$ is the hypergeometric function. The 
eigenenergies for the P{\"o}schl-Teller potential are \cite{flugge}
\begin{equation}
\label{ee}
E_n = {1 \over 2} V_0 (\alpha + \beta + 2n)^2
\end{equation}
From the shape of both the P{\"o}schl-Teller potential and the
effective potential in Eq. (\ref{effv}) this is exactly the kind of 
dependence one would expect for the energy eigenvalues. For small 
energies ({\it i.e.} $\alpha + \beta > 2n$) both potentials behave like 
a harmonic oscillator potential and so one would expect that the
leading term in $E_n$ should go like $2 V_0 (\alpha + \beta) n \propto
n$. For large energies ({\it i.e.} $2n > \alpha + \beta$) both
potentials behave like infinite spherical wells, and so one
would expect that the leading term in $E_n$ should go like
$2 V_0 n^2 \propto n^2$. 
As a simple example we will consider the $l=0$ case for the
potential in Eq. (\ref{effv}). For $l=0$ we find $J= 1/2$, $D=0$
$F=0$ and the potential in Eq. (\ref{effv}) becomes
\begin{equation}
\label{effv1}
V_{eff} (x) = {3 - 8x + 8x^2 \over 4 x^2 (1-x)^2}
\end{equation}
This potential approaches $3/ (4 (1-x)^2)$ as $x \rightarrow 1$ so
the test particle is just confined to the range $0<x<1$. In this 
range $V_{eff} (x)$ of Eq. (\ref{effv1}) reaches its minimum value
of $4$ at $x = 1/2$, and the potential is symmetric about this point.
In order for the P{\"o}schl-Teller potential 
to also be symmetric about $x=1/2$, and to also take
a value of $4$ at this point we can choose $V_0 = 1$ and
$\alpha = \beta = 2$. Now inserting these into Eq. (\ref{ee}) and
remembering that our eigenvalue from Eq. (\ref{kg6}) is $(E^2 -m^2)
/C^2$ we find that the approximate energy of the bound states for
this case with $l=0$ is
\begin{equation}
\label{ee1}
E^2 _n = m^2 + C^2 (2 + n)^2
\end{equation}
Note that this energy depends on the arbitrary constant $C$, which
sets the radius of the confining sphere ($r= 1/C$). As $C$ increases
the radius of the spherical shell decreases and from Eq. (\ref{ee1})
the energy of the state increases as would be expected.
Although in this $l=0$ case it was particularly easy to determine
$V_0, \alpha , \beta$, the form of the bound state energy given
by Eq. (\ref{ee1}) will be similar even when $l \ne 0$. 

\section{Discussion and Conclusions}

In this article we have presented a variety of solutions to
the field equations of Yang-Mills theory. Although finding exact
solutions to non-linear field theories is in general difficult,
many of the present solutions were found by using the mathematical
connection which exists between Yang-Mills theory and general 
relativity. Since general relativity has been studied for a longer
time than Yang-Mills theory there exists a body of known solutions
which can serve as guides for finding solutions to the Yang-Mills
or Yang-Mills-Higgs field equations. The Schwarzschild solution
of general relativity, both without and with a cosmological term,
gave rise to the solution with a spherical singularity in
Eq. (\ref{schsol}) and the linearly increasing solution of
Eq. (\ref{linear}). Although both of these solutions suffered
from the apparent drawback of having an infinite field energy,
they also exhibited some possible connection with the confinement
phenomenon. The linear solution of Eq. (\ref{linear}) is of the
form of phenomenological potentials \cite{eich} that are often 
used in studies of heavy quark bound states. In addition lattice
gauge theory arguments \cite{wilson} favour a linear type of
confining potential. The Schwarzschild-like solution of Eq.
(\ref{schsol}) has some similarities to bag models for
quark bound states. Spherical singularities, similar to those 
of the Schwarzschild-like solution, were also found to occur in
several other solutions as given in Eqs. (\ref{soln1}) (\ref{soln2}).
Actually, the solutions given in Eqs. (\ref{soln1}) (\ref{soln2})
possessed an infinite set of concentric spheres on which the gauge
and scalar fields became infinite. Thus, such spherically singular
surfaces may not be uncommon features of Yang-Mills field
theories. The SU(2) Schwarzschild-like solution can easily be generalized
to SU(N) \cite{sing4} by simply embedding the SU(2) 
solutions into an SU(N) gauge theory. It has also recently been found 
\cite{dzhunu} that true SU(3) solutions, which are not simply 
embeddings of the SU(2) solutions, can be given.

In the previous section the behaviour of a scalar test
particle placed inside the background potential presented by 
the Schwarzschild-like solution was examined. 
In order for the Schwarzschild-like potential to confine the test
particle, $\Phi ^A$, that it was necessary to couple, $\Phi ^A$,
to the scalar part of the Schwarzschild-like solution, $\phi ^a$,
via the coupling $m^2 \rightarrow (m + \lambda \sigma ^a \phi ^a /2)^2 $,
where $\lambda$ is the strength of the coupling between $\Phi ^A$
and $\phi ^a$. Even with this coupling it was found that confinement
occurred when $\lambda \ge g$, while for $\lambda < g$ there would be
some finite probability for $\Phi ^A$ to tunnel out of the spherical
well. Although normally it is thought that the confinement
phenomenon is the result of only gauge interactions, there has
been some work \cite{goebel} \cite {tekuda} \cite{ram} which indicates
that an effective scalar interaction may be needed to completely
explain confinement.  Another interesting aspect of the bound
state system studied in the previous section is that the total
system was a fermion even though only integer spin fields were involved.
The spin 1/2 nature of the bound state system resulted from the fact
that the isospin 1/2 of the test particle $\Phi ^A$ was converted
into spin 1/2 when it was placed inside the Schwarzschild-like
solution. Another way of arriving at this result is to note that almost
all of the solutions presented here could be shown to carry a
topological magnetic charge. Thus, in the same way that a standard magnetic
charge - electric charge system carries a field angular momentum
of $1/2$ in their combined electromagnetic fields, so too the
combined charges of the Schwarzscild-like solution and $\Phi ^A$
carried a field angular momentum of $1/2$ in their combined
non-Abelian fields. If a realistic model of hadronic bound
states can be constructed from these classical field theory solutions,
then the fact that the net angular momentum of these states does not
come entirely from the constituent particles, may offer a possible 
explanation of the EMC effect \cite{ashman}, which shows that a large 
part of the net spin of the proton does not come from the valence
quarks.

In addition to the Schwarzschild-like solutions presented
here it is also possible to take more complex solutions
from general relativity to find other  Yang-Mills solutions.
In Ref. \cite{sing6} the general relativistic Kerr solution
was used to construct a new Yang-Mills solution. Although the
final form of the Yang-Mills Kerr-like solution was not as
simple as the Schwarzschild-like solutions, it did share the
common feature of having confining surfaces on which the
fields became singular. Finally it is also possible to use
this method for finding solutions to non-linear field
equations in reverse : starting from known solutions to
the Yang-Mills equations one can obtain solutions to the
general relativistic field equations \cite{sing7}.

\section{Dedication} This article is dedicated to the memory
of Professor Fyodor Lunev.

\end{document}